\documentclass[sn-nature]{sn-jnl} 

\usepackage{graphicx}%
\usepackage{multirow}%
\usepackage{amsmath,amssymb,amsfonts}%
\usepackage{amsthm}%
\usepackage[title]{appendix}%
\usepackage{xcolor}%
\usepackage{textcomp}%
\usepackage{manyfoot}%
\usepackage{booktabs}%
\usepackage{algorithm}%
\usepackage{algorithmicx}%
\usepackage{algpseudocode}%
\usepackage{listings}%
\usepackage{upgreek}
\usepackage{soul,color}
\usepackage{fix-cm}
\usepackage[version=4]{mhchem}

\begin{document}

\title[]{All-optical stochastic switching of magnetization textures in \ce{Fe3Sn2}}

\author*[1,2]{\fnm{Jonathan T.}\sur{Weber}}
\email{jonathan.weber@physik.uni-regensburg.de} \equalcont{These authors contributed equally to this work.}
\author*[3]{\fnm{Andr\'as} \sur{Kov\'acs}} 
\email{a.kovacs@fz-juelich.de} \equalcont{These authors contributed equally to this work.}
\author*[4]{\fnm{Michalis} \sur{Charilaou}}
\email{michalis.charilaou@louisiana.edu}
\author[3,5]{\fnm{Deli} \sur{Kong}}
\author[6]{\fnm{Lilian} \sur{Prodan}}
\author[6,7]{\fnm{Vladimir} \sur{Tsurkan}}
\author[1]{\fnm{Alexander}\sur{Schr\"oder}}
\author[8]{\fnm{Nikolai S.} \sur{Kiselev}}
\author[6]{\fnm{Istv\'an} \sur{K\'ezsm\'arki}}
\author[3]{\fnm{Rafal E.} \sur{Dunin-Borkowski}}
\author*[3]{\fnm{Amir H.}\sur{Tavabi}}
\email{a.tavabi@fz-juelich.de}
\author*[1,9]{\fnm{Sascha}\sur{Sch\"afer}} 
\email{sascha.schaefer@physik.uni-regensburg.de}

\affil[1]{\orgdiv{Department of Physics}, \orgname{University of Regensburg},  \city{Regensburg}, \postcode{93053}, \country{Germany}}

\affil[2]{\orgdiv{Institute of Physics}, \orgname{University of Oldenburg},  \city{Oldenburg}, \postcode{26129}, \country{Germany}}

\affil[3]{\orgdiv{Ernst Ruska-Centre for Microscopy and Spectroscopy with Electrons}, \orgname{Forschungszentrum J\"ulich},  \city{J\"ulich}, \postcode{52425}, \country{Germany}}

\affil[4]{\orgdiv{Department of Physics}, \orgname{University of Louisiana at Lafayette},  \city{Lafayette, Louisiana}, \postcode{70504}, \country{USA}}

\affil[5]{\orgdiv{Department of Materials Science and Engineering}, \orgname{Southern University of Science and Technology},  \city{Shenzhen}, \postcode{518055}, \country{China}}

\affil[6]{\orgdiv{Experimental Physics V}, \orgname{University of Augsburg},  \city{Augsburg}, \postcode{86135}, \country{Germany}}

\affil[7]{\orgdiv{Institute for Applied Physics}, \orgname{Moldova State University},  \city{Chisinau}, \postcode{MD 2028}, \country{R. Moldova}}

\affil[8]{\orgdiv{Peter Gr\"unberg Institute and Institute for Advanced Simulations} \orgname{Forschungszentrum J\"ulich}, \city{J\"ulich}, \postcode{52425}, \country{Germany}}

\affil[9]{\orgdiv{Regensburg Center for Ultrafast Nanoscopy}, \orgname{University of Regensburg},  \city{Regensburg}, \postcode{93040}, \country{Germany}}

\abstract{
The all-optical control of magnetization at room temperature broadens the scope of applications of spin degrees-of-freedom in data storage, spintronics, and quantum computing. Topological magnetic spin structures, such as skyrmions, are of particular interest due to their particle-like properties, small size and inherent stability. Controlling skyrmion states without strong magnetic fields or large current densities would create new possibilities for their application. In this work, we utilize femtosecond optical pulses to alter the helicity of the spin configuration in dipolar skyrmions formed in the kagome magnet \ce{Fe3Sn2} in the absence of an external magnetic field and at room temperature. \emph{In situ} Lorentz transmission electron microscopy is used to visualize the stochastic, light-induced switching process of chiral N\'eel caps, while the internal Bloch component of the dipolar skyrmions remain unchanged. In addition to this switching process, we observe the interconversion between type I skyrmionic and type II bubble configurations depending on the external magnetic field and illumination conditions. To corroborate the spin states and the light-induced magnetization dynamics, micromagnetic modelling and simulations of the resulting electron phase shift maps are conducted to elucidate the spin rearrangement induced by individual femtosecond optical pulses. 
}


\maketitle

\section{Introduction}
Magnetic skyrmions and hopfions \cite{Tokura2020, Zheng2018, Kiselev2023} are localized topological swirling spin field configurations that have gained significant research interest over the past decade because of their nanometric scale and unique properties, which make them ideal information carriers for high-density fast data storage and for information processing applications based on neuromorphic and stochastic computing schemes \cite{kaiser_probabilistic_2021}.
Since the first experimental detection of skyrmions in the chiral magnet \ce{MnSi} \cite{muhlbauer_skyrmion_2009} and their real-space observation by Lorentz transmission electron microscopy (TEM) shortly after \cite{yu_near_2011}, several skyrmion-hosting bulk and thin film heterostructures have been discovered with various electric and magnetic properties such as topological Hall effects and ultra-low current density control of their motion \cite{Fert2017, Du2022}. Besides the different systems possessing bulk or interfacial Dzyaloshinskii-Moriya interaction to stabilize skyrmions, layered kagome magnets, such as \ce{Fe3Sn2} and \ce{Co3Sn2S2}, offer an interesting alternative for hosting particle-like spin textures. The rhombohedral material \ce{Fe3Sn2} has received increasing interest due to its intriguing properties such as a large dc and optical anomalous Hall effect \cite{Fenner2009,Kida_2011, Schilbert2022}, flat electron bands near the Fermi energy \cite{Lin2018}, and massive Dirac fermions \cite{Ye2018, Kang2020}, which are combined with a high Curie temperature (670 K) \cite{Fenner2009}. 
\ce{Fe3Sn2} is a topological kagome magnet, in which the Fe and Fe-Sn bilayers are stacked along the crystallographic c-axis in corner-sharing kagome triangles. The competing magnetic uniaxial ($K_u$) and shape anisotropies in thin film samples of this material \cite{Kezsmarki2021} lead to unconventional topological spin textures defined as dipolar skyrmions. In such particles, the spin texture consists of a Bloch-type spin arrangement in the mid-section of the film and two N\'eel-type surface spin twists which can have parallel or opposite helicities \cite{Kong2023b, Du2024}. 
Recently, the field-driven generation of dipolar skyrmions in \ce{Fe3Sn2} and their simultaneous detection by \emph{in situ} Lorentz TEM and the anisotropic magnetoresistance has been successfully demonstrated \cite{Du2023}.

To achieve low-power and fast control of relevant degrees-of-freedom in topological spin textures, various processes have been proposed including magnetic, electric, thermal and light stimuli \cite{klaui_perspective_2018}. 
In the field of information recording technology \cite{Rasing2025}, the optical control of magnetic states \cite{Lambert2014} has emerged as a particularly promising direction, extending beyond the realm of heat-assisted magnetic recording (HAMR). This approach holds considerable appeal due to its reliance on low-power requirements for optical pulse generation, the capability to focus optical probes down to sub-micrometer dimensions, the absence of external magnetic fields, and the potential for ultrafast magnetic switching speeds. 
The ultrafast magnetization changes in ferromagnetic or antiferromagnetic samples are normally achieved on femto or picosecond (fs or ps) light pulse excitation, where the optical pulses can demagnetize a localized region followed by thermal quenching \cite{beaurepaire_ultrafast_1996, Rasing2010, Schafer2017, Adam2025}. 
Beyond the ultrafast demagnetization of ferromagnetic materials, also the all-optical switching of the magnetization direction in ferrimagnets has been demonstrated, revealing intriguing non-equilibrium dynamics \cite{stanciu_all-optical_2007,radu_transient_2011}.

While there has been tremendous progress in fast and ultrafast Lorentz TEM \cite{park_4d_2010, schliep_picosecond_2017, Schafer2018, Carbone2018, Ropers2020, Zhu2024, liu_correlated_2025, fan_spatiotemporal_2025}, a detailed understanding of the light-induced manipulation of spin textures can also be achieved by \emph{in situ} Lorentz TEM studies with femtosecond optical excitation \cite{Schafer2017, Zhu2018, Fu2024}. Using Lorentz TEM, light-pulse-induced effects on skyrmion-hosting materials have been demonstrated in the chiral B20-type helimagnet FeGe \cite{Carbone2018}, in the chiral Co-Zn-Mn alloy system \cite{Shimojima2021}, in the Mott insulator \ce{Cu2OSeO3} \cite{Carbone2022} and in 2-dimensional van der Waals-type \ce{Fe_{3-x}GaTe2} \cite{Fu2024}. Whereas previous studies have focused on the creation and annihilation of topological spin textures as well as on the control of their motion, an optical control of internal degrees-of-freedom has remained largely elusive. In a recent study, the ultrafast optical control of the collective breathing mode of a skyrmion ensemble was demonstrated \cite{Carbone2022} but without remanent changes to individual skyrmions. Considering possible applications of topological spin structures in the fields of information storage and neuromorphic computation, it would be highly beneficial to encode information not in the presence or absence of these topological structures but instead in stable internal states accessible to outside stimuli. 

In this work, we demonstrate all-optical, stochastic magnetization modulation of a topological spin structure at room temperature. Specifically, we show that the magnetization winding of N\'eel spin twists in dipolar skyrmions within kagome-type \ce{Fe3Sn2} thin films can be changed by femtosecond optical light pulses. The ultrafast switching phenomena were monitored by \emph{in situ} Lorentz TEM imaging at magnetic remanence and at room temperature. A statistical evaluation of the electron micrographs uncovered a stochastic mechanism of spin texture variation within the optically excited dipolar skyrmions, while retaining the Bloch-type internal circular domain wall chirality. In addition, utilising an off-axis magnetic field as an additional experimental degree-of-freedom, the optically induced interconversion of type II bubbles and type I skyrmions was observed, involving a substantial spin reorganization in the Bloch- and N\'eel structures. Theoretical calculations, based on micromagnetic modelling and simulations of the electron phase shift, provide insights into the magnetization dynamics of the dipolar skyrmions induced by the ultrafast light pulse. The kagome-type \ce{Fe3Sn2} material thus constitutes an intriguing platform for the development of all-optical magnetization switching of topological particles.

\section{Results and Discussion}

\subsection{Spin structure in dipolar skyrmions}

\begin{figure*}[ht]%
\centering
\includegraphics[width=1.0\textwidth]{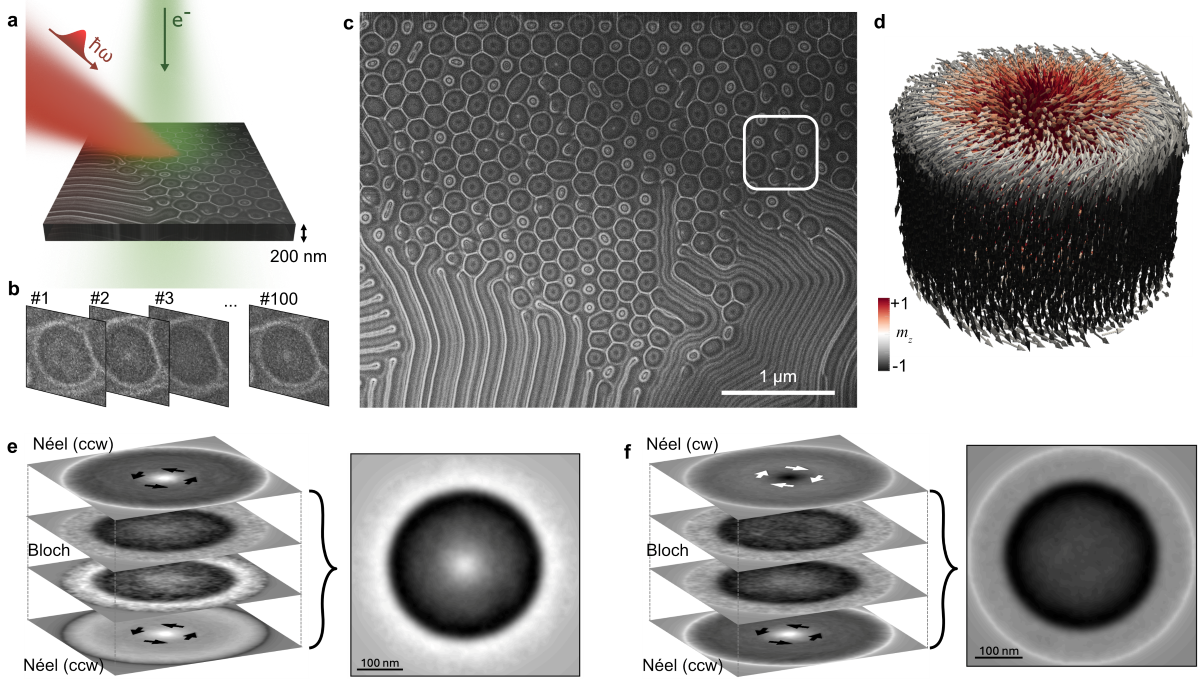}
\caption{The TEM experiment and magnetic state of dipolar skyrmions in \ce{Fe3Sn2}.
\textbf{a,b} Scheme of the experimental setup. Series of Lorentz TEM images are recorded for 1 s following a 169-fs light pulse.
\textbf{c} Lorentz TEM micrograph of magnetic states in \ce{Fe3Sn2} lamella recorded at magnetic remanence. An overfocus of 1.9~mm was used. The rectangle marks a cluster of type II magnetic bubbles.
\textbf{d} Micromagnetic simulation of a type I dipolar skyrmion formed in a disc-shaped sample. 
\textbf{e,f} Segmented phase shift imparted on an electron beam by a dipolar skyrmion with (e) matching and (f) opposing field rotations of the N\'eel states. The Bloch domain state in the sample core is the same in both cases. The total phase shift (left panels) is calculated by summation over the segmented contributions.}
\label{fig1}
\end{figure*}

For investigating the optical response of \ce{Fe3Sn2} by \emph{in situ} Lorentz TEM (Fig.~\ref{fig1}a,b), a single-crystalline, electron-transparent lamella was prepared (for details see Methods), which exhibits a stripe-like domain pattern in the magnetic ground state with alternating out-of-plane magnetic field alignment \cite{tang_two-dimensional_2021}.
By applying an external magnetic field of 0.53~T perpendicular to the kagome plane (magnetic easy axis), dipolar skyrmions were generated and subsequently imaged at magnetic remanence. A typical Lorentz TEM micrograph recorded with 1.2~mm image defocus is presented in Fig.~\ref{fig1}c, showing regions characterized by stripe-like magnetic domain states alongside numerous circular-shaped particles displaying different contrast patterns. The majority of the particles have been identified as dipolar skyrmions of type I, which can be described as a cylindrical domain with a magnetisation antiparallel to the surrounding out-of-plane spin orientations, and continuous closure domain walls. The mid section of the dipolar skyrmions shows a Bloch-type texture with clockwise (cw) or counter-clockwise (ccw) field rotation, while close to the surfaces the field arrangement aligns in a twisted N\'eel structure. Since type I bubbles can be described with the same integer topological charge \emph{Q} as chiral Bloch skyrmions, they are termed dipolar skyrmions. As marked by a rectangle in Figure \ref{fig1}c, some particles show contrast variations in their circular shape, indicating domain walls with a spin alignment oriented towards the in-plane direction. These particles are identified as type II magnetic bubbles, which have a vanishing topological charge \cite{Kong2023b}. 

We observe different helicities of the type I dipolar skyrmions in \ce{Fe3Sn2}, characterized by either bright or dark Lorentz contrast. The dipolar skyrmions have a hybrid magnetic state containing both Bloch- and N\'eel-type magnetic textures with a smooth transition, moving from the mid-section of the film to the top and bottom surfaces. Whereas the helicity of the Bloch domain wall determines the Lorentz contrast in the outer parts of the dipolar skyrmion, the presence of twisted N\'eel caps and their respective helicities is the contributing factor to the spot-like contrast observed in the middle of dipolar skyrmions and type II bubbles. 
Figure~\ref{fig1}d shows the spin texture of a type I dipolar skyrmion, obtained by a micromagnetic simulation (for details see Methods and Section~\ref{umag}). The hybrid Bloch-N\'eel structure generates a characteristic Fresnel image contrast sensitive to in-plane components of the magnetic field. Specifically, a bright or dark central dot appears when the rotation directions of the bottom and top N\'eel caps are both ccw or cw, respectively. In cases where the upper and lower surfaces exhibit opposite spin rotation directions, the central image intensity is neutral. All three cases are also visible in the Lorentz micrograph displayed in Figure~\ref{fig1}c. To further disentangle the individual contributions of the different parts of the dipolar skyrmion on the total Lorentz contrast, we calculated the electron phase shift maps for thin film slices for the case of equal N\'eel cap rotation at both surfaces, as shown in Figure~\ref{fig1}e. As visible in the slices, the image contrast in the center of the skrymion originates from both surfaces, whereas the outer Bloch-type contrast is dominated from mid-section slices. For the co-rotating N\'eel caps, the contribution from both surface add up in the total phase shift Figure~\ref{fig1}e (right panel). In contrast, for counter-rotating N\'eel caps (Figure~\ref{fig1}f) the surface contributions cancel out, leaving a neutral image contrast in the center of the dipolar skyrmion (see Supplementary Fig.~S1 for further combinations of Bloch wall and N\'eel cap helicities). \\

\subsection{Light-induced switching processes }

\begin{figure*}[ht]%
\centering
\includegraphics[width=1.0\textwidth]{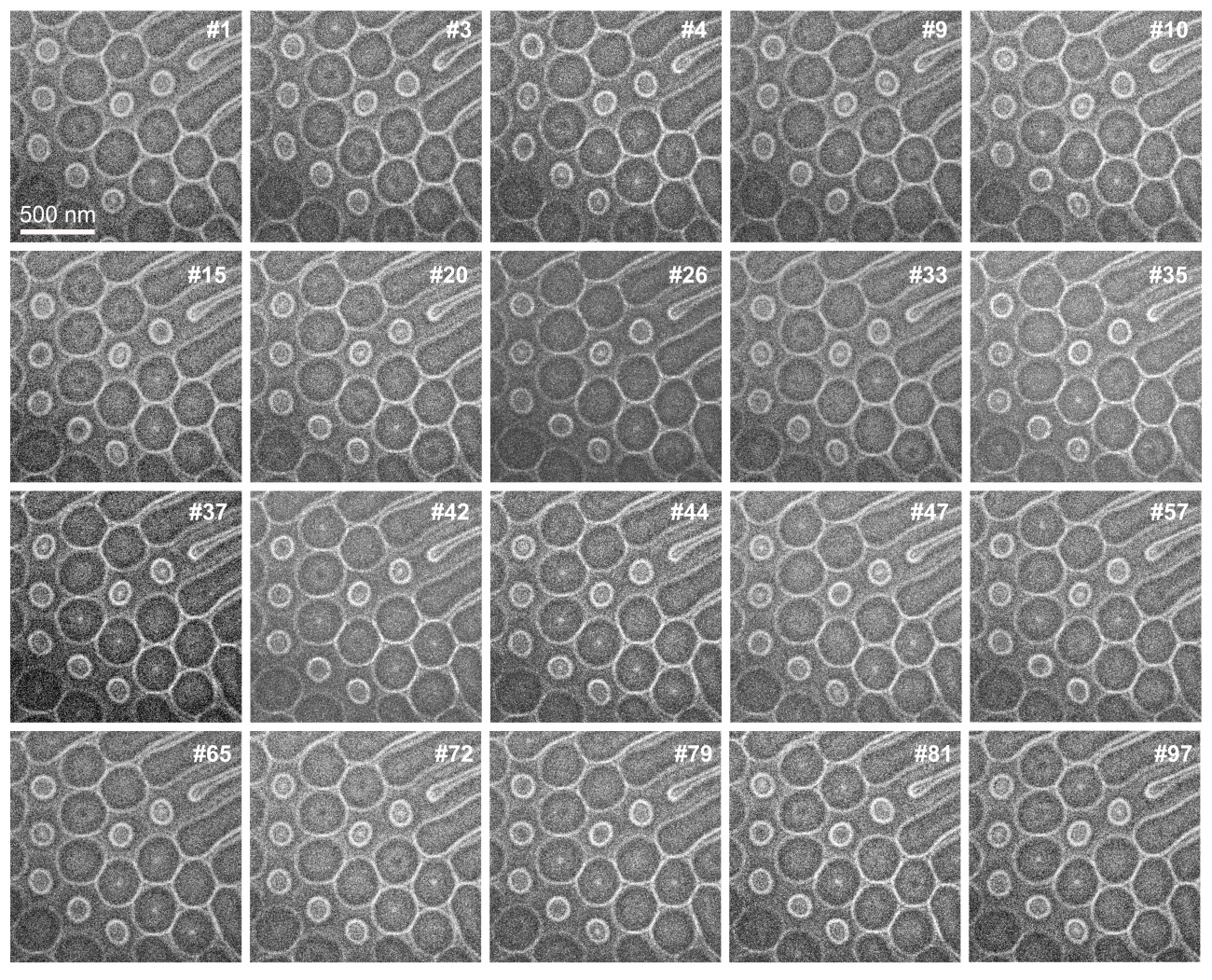}
\caption{Sequence of Lorentz TEM micrographs of dipolar skyrmions in \ce{Fe3Sn2} recorded after applying femtosecond light pulses at magnetic remanence. The number of applied pulse is indicated in each inset. Note the variation of the central contrast within the circular image features showing either a bright or dark dot or a constant contrast. Optical pulse energy: 100 nJ.} 
\label{fig2}
\end{figure*}

As an experimental platform to study the optical control of internal degrees-of-freedom in dipolar skyrmions, we utilize the ultrafast transmission electron microscope (UTEM) at the Regensburg Center for Ultrafast Nanoscopy (RUN).
Upon illumination of the \ce{Fe3Sn2} lamella by a single, femtosecond optical pulse, as depicted in Fig.~1a (800~nm central wavelength, 169~fs pulse duration, p-polarized), Lorentz TEM images in the equilibrium state and in magnetic remanence were recorded. A series of such micrographs is presented in Figure~\ref{fig2}. After each light pulse, we observe slight variations in the position and shape of individual particles, and, more strikingly, significant contrast fluctuations are observed at the centres of dipolar skyrmions, where the image contrast can invert, diminish, or reappear, showing either bright or dark contrast. The outer ring of the circular contrast features, however, remains unaffected by the optical excitation. The observed phenomenon is attributed to a light-induced switching of the helicity of the N\'eel caps of individual dipolar skyrmions, while the core Bloch domain wall maintains its configuration. An assembled video of the \ce{Fe3Sn2} specimen subjected to 100 consecutive incident optical pulses, each with an energy of 100~nJ, is provided as Supplementary Movie 1.

In our experiments, the incident optical pulse energy was varied between 75~nJ and 125~nJ, which does not seem to have a significant effect on the core contrast variations (see Supplementary information). In the limit of low pulse energies (below 75~nJ), a substantial decline in switching frequency was found. Conversely, for pulse energies exceeding 250~nJ, structural damage and non-reversible magnetic contrast changes were observed. 

\begin{figure*}[ht]%
\centering
\includegraphics[width=1.0\textwidth]{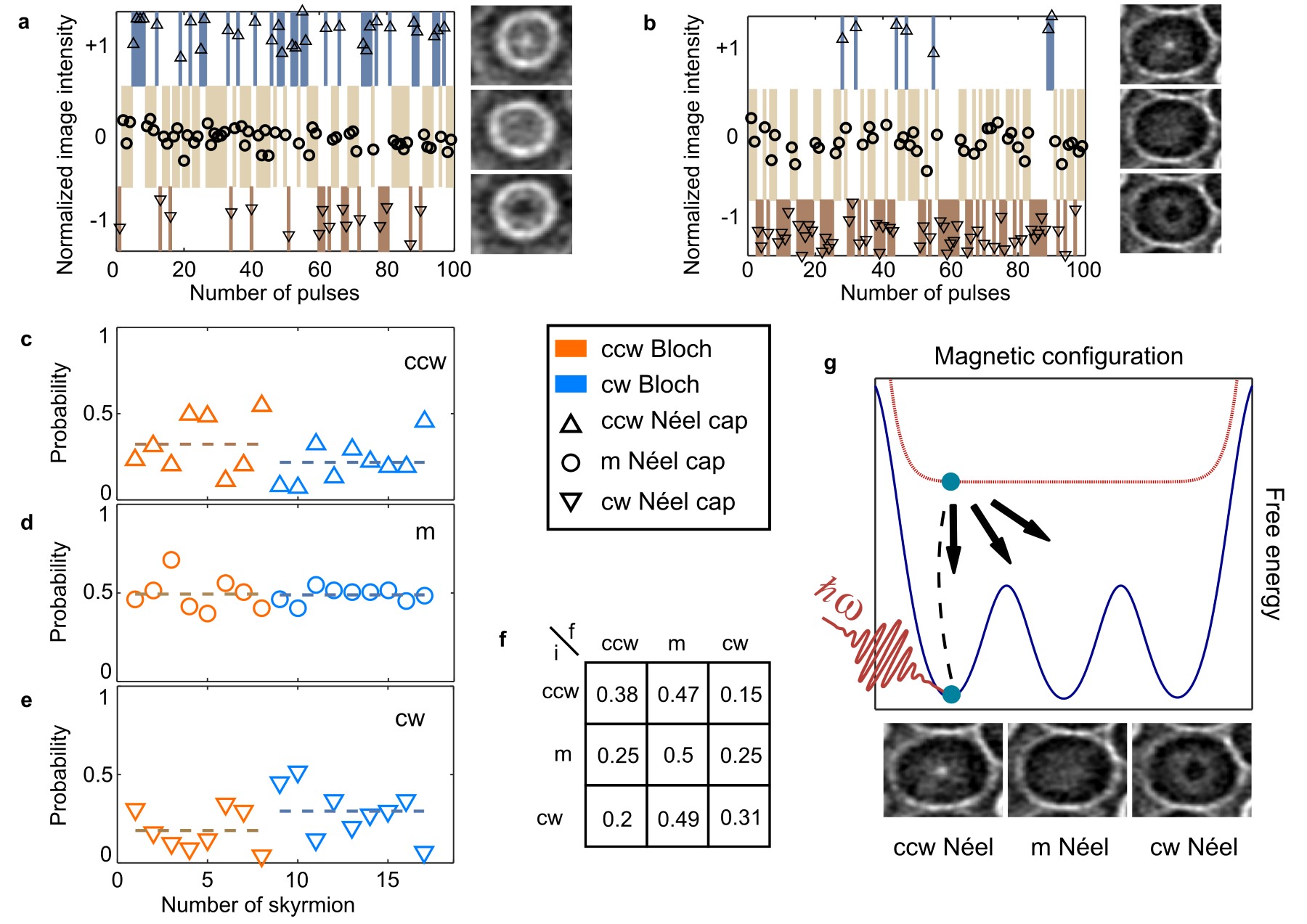}
\caption{Statistical analysis of light-induced N\'eel cap magnetisation switching behaviour in \ce{Fe3Sn2}. 
\textbf{a,b} Normalized image intensity in a circular region within the centre of one dipolar skyrmion with (a) ccw and (b) cw Bloch core rotation. The coloured boxes indicate the threshold values used to discriminate between clockwise/mixed/counter-clockwise rotation states ($\nabla$/o/$\Delta$) of the N\'eel caps. The Lorentz TEM images on the right exemplify the three observed states. The images were Fourier filtered to enhance the visibility. Incident optical pulse energy: 100~nJ; measurements were performed at magnetic remanence.
\textbf{c-e} Probability of the N\'eel cap to be found in the (c) ccw, (d) m or (e) cw state. The data was averaged over all 100 incident optical pulses for a total of 16 analysed dipolar skyrmions. Brown (blue) symbols indicate data for skyrmions with a cw (ccw) rotation of the Bloch domain wall.
\textbf{f} Probability matrix for the transitions from the initial N\'eel cap magnetic state (i) to the final state (f). Probabilities are averaged over 100 incident optical pulses and 16 analysed dipolar skyrmions.
\textbf{g} Sketch of the free energy landscape for a dipolar skyrmion with femtosecond optical excitation. From the excited hot-spin state, the system falls into one of the local minima of the free energy with the probability described by the matrix in (f).}
\label{fig3}
\end{figure*}

In order to extract correlations within the stochastic light-induced switching process, subsequent Lorentz TEM images were analysed, each recorded after excitation with a single optical pulse for a total of up to 100 incident pulses. 
Histograms of the distribution of the image intensity within a circular region of the dipolar skyrmions show three distinct peaks (see Supplementary information) allowing for a classification of the N\'eel caps into one of three states: the bright contrast state, corresponding to ccw rotating N\'eel caps at the top and bottom of the Fe$_3$Sn$_2$ lamella (ccw state); the mixed state, associated with opposite helicities of the two N\'eel caps (m state); or the dark contrast state, representing a cw rotation of the N\'eel caps (cw state). 
To clearly distinguish between these states, upper and lower image contrast threshold values were assigned to each individual dipolar skyrmion based on the minima of the image intensity distribution histograms (see also Supplementary information). 
Figure~\ref{fig3}(a,b) presents the results of two selected analyses for two dipolar skyrmion with ccw and cw rotation of the Bloch wall, respectively.
By averaging over all incident pulses, the probability of an individual dipolar skyrmion adopting the ccw, m, or cw state after excitation by a single light pulse is determined. These probabilities are shown in Fig.~\ref{fig3}c-d for a total of 16 dipolar skyrmions. Notably, the N\'eel caps of all investigated dipolar skyrmions adopt the mixed state in approximately 50~\% of the events, with relatively small variation (m state standard deviation (std): 0.07) compared to the inter-skyrmion variation observed for the other states (ccw state std: 0.14, cw state std: 0.13). Furthermore, it has been observed that dipolar skyrmions with a ccw Bloch wall exhibit a slightly higher probability for the N\'eel caps to adopt the ccw state, while the reverse is true for skyrmions with a cw Bloch wall. However, this asymmetry is relatively minor compared to the fluctuations observed between individual skyrmions. Measurements conducted at varying incident optical pulse energies reveal an overall similar switching behaviour, with a slightly enhanced asymmetry in the probabilities for ccw and cw Bloch walls at incident pulse energies of 125~nJ (see the Supplementary information).
This small probability asymmetry is tentatively ascribed to a magnetic coupling between the rotation of the N\'eel caps and the internal Bloch wall.
However, the substantial inter-skyrmion variations imply that the micromagnetic environment of the dipolar skyrmion exerts a more significant influence on the probability of the N\'eel caps adopting a particular magnetic state.
The switching probabilities from an initial state (i) to a final state (f) after excitation with a single light pulse with a pulse energy of 100~nJ is presented as a probability matrix in Fig.~\ref{fig3}f (averaged over dipolar skyrmions with ccw and cw Bloch walls). 
It is hypothesised that optical excitation will lead to an increased spin temperature, levelling out the corrugation of the spin-free-energy landscape and allowing for an interconversion of individual N\'eel cap states, as illustrated in Fig.~\ref{fig3}g. As the system cools by heat transfer to the environment, the magnetic configuration relaxes into one of the local minima of the free magnetic energy landscape, a process potentially influenced by the local magnetic environment. If the N\'eel cap configuration is adopted randomly, the final N\'eel cap state should be independent of the initial state and thus all rows of the probability matrix should be equal. Given that the mixed state can be formed by two N\'eel cap configurations, the rows of the model probability matrix are given by $\left[ \frac{1}{4}~\frac{1}{2}~\frac{1}{4} \right]$.
In contrast, the experimentally determined probability matrix demonstrates a dependence on the initial state and only the switching from the mixed state proceeds with the expected probabilities of a random process. Starting from either the ccw or cw twisted N\'eel state, the dipolar skyrmion tends to remain in the original configuration, demonstrating that the simultaneous reversal of both N\'eel cap helicities is less likely. At the same time, the probability of switching to the mixed state remains approximately 50~\%, indicating that at the experimentally adopted optical excitation level, the helicity of one N\'eel cap is randomly selected. \\
Notably, applying a magnetic field perpendicular to the sample surface using the electron microscope's objective lens does not alter the overall switching behaviour but significantly enhances the previously mentioned asymmetry (see Supplementary information for data acquired under an external field of 133~mT). Potential approaches for controlling the N\'eel cap switching behaviour by fine-tuning the external magnetic field are currently under investigation.

\subsection{Interconversion of type I and type II structures}

\begin{figure*}[ht]%
\centering
\includegraphics[width=0.9\textwidth]{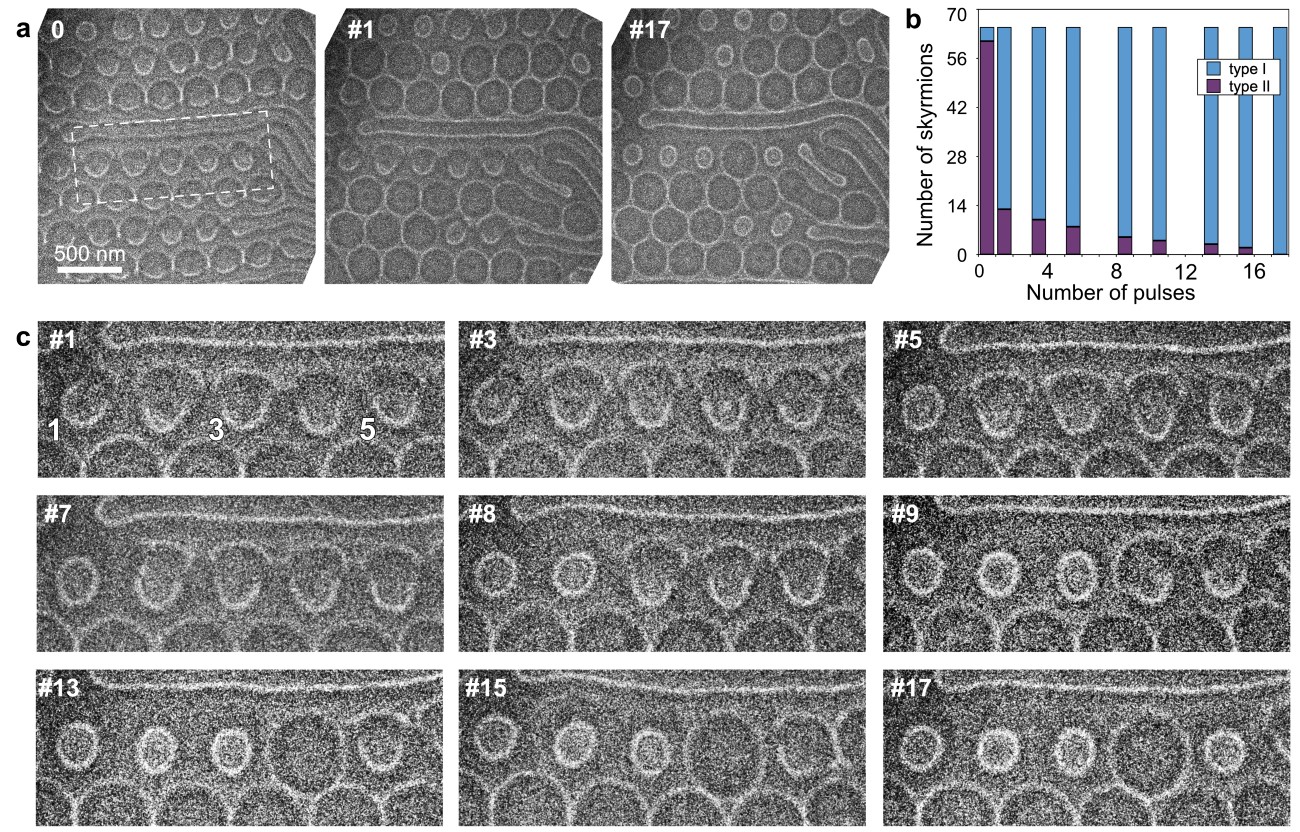}
\caption{Light-triggered transition of type II magnetic bubbles to type I dipolar skyrmions using a light pulse energy of 75~nJ at magnetic remanence.
\textbf{a}~Lorentz TEM images showing the magnetic structures before light illumination (0), and after shot \#1 and \#17, respectively. 
\textbf{b}~Number of type I and type II structures changes after 17 pulses.
\textbf{c}~Lorentz TEM images showing snapshots of the transition from type II to type I. Upper left corners the number of the light-pulse is indicated.}
\label{fig4}
\end{figure*}

Following the implementation of a protocol involving the application of an external magnetic field of 0.4~T to the slightly tilted \ce{Fe3Sn2} lamella, the predominant type of produced magnetic particles was found to be of a type II structure. In particular, Lorentz TEM images recorded at magnetic remanence reveal the presence of Bloch wall features with alternating contrast (Fig.~\ref{fig4}a). In the analysed image region, 65 particles were present, 4 of which had a type I and 61 had type II structure. After 17 optical pulses with a pulse energy of 75~nJ, a transition to a type I was observed in all type II bubbles. The number of particles of the two types observed after each light pulse is plotted in Fig.~\ref{fig4}b. Following the conversion of all bubbles from type II to type I, the newly formed dipolar skyrmions exhibit 43 ccw and 22 cw Bloch magnetization rotations. Figure~\ref{fig4}c illustrates the step-by-step changes of five type II bubbles which exhibit uniform Bloch helicities. In particle 1, the segment between its two internal domain walls shortened after shots \#2 and \#4, forming a type I structure with ccw field rotation after shot \#5. After shot \#9, three out of five particles were transformed into type I structures with the same Bloch helicity. Interestingly, particle number 4, formed after shot \#12, displays a cw Bloch rotation despite having the same initial internal state and similar magnetic environment as the others. This observation suggests that the light-triggered type II to type I transition is, at least in part, a random process, and is likely connected to the mobility of the Bloch domain defects and their annihilation dynamics in the partially melted spin configuration shortly after the optical pulse.
In the absence of an applied field, only the interconversion from type II to type I structures was observed, but not in the opposite direction. This finding suggests that the type II structure possesses a higher free energy as compared to the type I bubbles, and that the activation free energy barrier in between these two spin configurations can be surpassed in the hot-spin state. Notably, at an applied field of approximately 133~mT, rare evidence for the light-induced generation of a type II bubble from a type I Bloch structure was observed at an incident optical pulse energy of 100~nJ (see Supplementary Movie M2). 

\subsection{Micromagnetic modelling}\label{umag}

\begin{figure*}[ht]%
\centering
\includegraphics[width=1.0\textwidth]{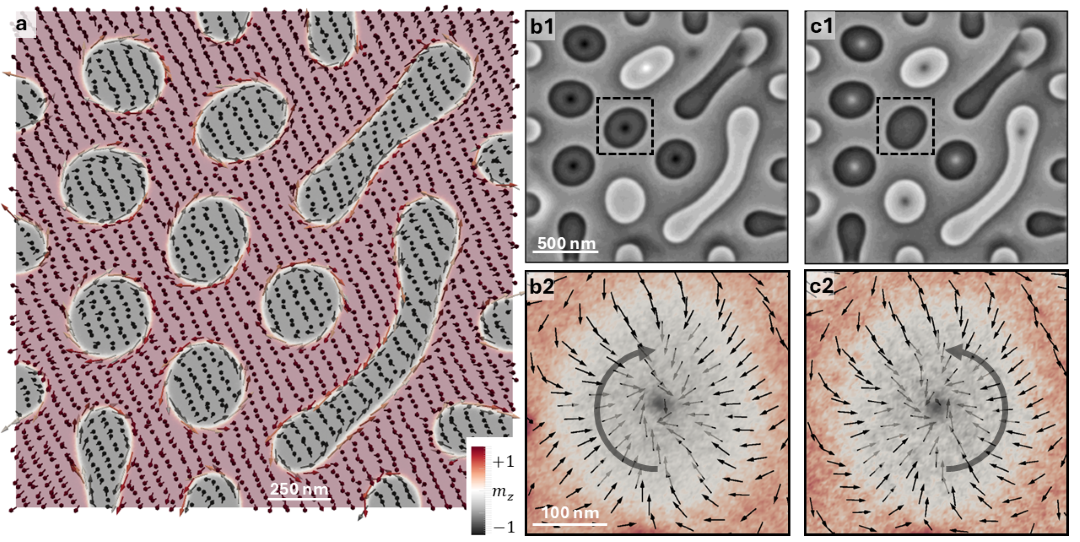}
\caption{Snapshots of micromagnetic simulations of a \ce{Fe3Sn2} lamella showing (a) a top-view of the cross section and calculated phase-shift maps (b1) before and (c1) after the temperature spike. The selected dipolar skyrmions indicated with dashed squares in the maps are shown in (c1) and (c2), respectively. The vanishing of the central dark dot contrast is due the change of the configuration of the top surface. }
\label{fig:MM}
\end{figure*}

To understand the fundamental processes involved in the formation of the dipolar skyrmion spin texture and the switching induced by femtosecond light pulses, micromagnetic simulations were performed based on the Landau-Lifshitz-Gilbert (LLG) equation (see Methods) to reveal details of the underlying spin texture variations. Taking values for the saturation magnetization and the uniaxial magnetocrystalline anisotropy from magnetometry experiments, we performed detailed micromagnetic simulations. For a direct comparison between experiment and theory, in particular with respect to long-range dipolar interactions, the simulation geometry was chosen to be comparable to the sample used in experiments: a thin film with dimensions of 2000~nm $\times$ 2000~nm $\times$ 250~nm, as shown in Fig.~\ref{fig:MM}. For the top and bottom surfaces of the film we considered soft ferromagnetic properties, in order to approximate the non-ideal structure resulting from the ion milling fabrication and possible oxidation of the TEM specimen. For the soft layers, the saturation magnetization and exchange stiffness were assumed to be the same as in the ideal case but with a vanishing crystalline anisotropy ($K_1=K_2=0$). Additionally, thin layers without modified surfaces were tested as well in order to investigate the magnetic state in the ideal case.

The influence of the femtosecond optical pulse was simulated as a temperature spike in the electron system, via a random temperature-dependent thermal field, by raising $T$ to a peak value within 20~ps and subsequently gradually lowering it to room temperature within 100~ps. Various values of $T_\mathrm{max}$ were tested to determine the effective peak temperature that best reproduces the experimental observations, and this was found to be $T_\mathrm{max}=12,000$ K. In principle, the relevant temperature in the experiments corresponds to the temperature of the spin system, which in metals is expected to increase after femtosecond optical excitation on a few-100-fs time-scale, resulting from electron-spin coupling. Due to the simultaneously occurring electron-phonon coupling, we expect a peak spin temperature on the order of several hundred Kelvin. However, in the micromagnetic simulation, macrospins are considered, neglecting spin fluctuations within a simulation cell which become important at high spin temperatures. Therefore, larger effective temperatures are required in the simulations to induce comparable spin dynamics as observed in the experiments.

Figure~\ref{fig:MM}a shows the top view of the simulated spin field in the thin-film mid-section depicting the Bloch-type domain wall surrounding each magnetic particle. Based on the three-dimensional spin field, we simulated the expected electron phase shift, as shown in Fig.~\ref{fig:MM}b1. All relevant magnetic features observed in the experiments, such as type I dipolar skyrmions, elongated bubble shapes, cw and ccw Bloch helicities, and the presence of N\'eel caps, are well reproduced in the simulation. The correspondingly simulated magnetic structure after applying a temperature peak and subsequent relaxation is presented in Fig.~\ref{fig:MM}c1.
Comparing the magnetic structure before and after excitation (Fig.~\ref{fig:MM}b1,c1), the experimentally observed switching of N\'eel caps can be clearly reproduced in the simulations. 
For example, the magnetic texture for the dipolar skyrmion highlighted in Fig.~\ref{fig:MM}b1,c1 changes at the top surface from cw to a ccw rotating N\'eel cap (Fig.~\ref{fig:MM}b2,c2). Together with the non-changing cw rotating N\'eel cap at the bottom surface (not shown), the contrast within the centre of the dipolar skyrmion switches from dark to neutral.

In the simulations, we have observed four possible scenarios for different $T_\mathrm{max}$ (not shown here). In the low-temperature limit, the pulse introduces a small degree of randomization but the micromagnetic state remains overall unchanged during and after the pulse. In the temperature range $11,000$ K $< T_\mathrm{max} < 18,000$ K, the temperature pulse provides enough thermal energy to randomly reverse the twisting of the N\'eel caps, while the body of the Bloch dipolar skyrmions remains unchanged. In the range $18,000$ K $< T_\mathrm{max} < 24,000$~K, the thermal pulse stochastically switches the configuration of both the N\'eel caps and the Bloch dipolar skyrmions. In the high-temperature limit, the Zeeman energy of the random field overcomes the ferromagnetic exchange, leading to a completely demagnetized state that does not return to its original configuration with dipolar skyrmions after the pulse is removed. 
The excellent agreement between simulation and experimental observation provides substantial support for our interpretation of the light-induced processes and captures the stochastic nature and diversity of the involved switching phenomena. 

\section{Conclusion}
Topological spin textures are widely regarded as promising candidates for next-generation high-density magnetic recording and non-conventional computation schemes owing to their nanometric scale, intrinsic stability and susceptibility to external stimuli such as current-induced spin-orbit or spin-transfer torques. In addition, the all-optical manipulation of these structures has emerged as an alternative approach, offering the potential for reduced power consumption and ultrafast versatility. However, the necessity for a magnetic field to generate and stabilize skyrmion phases, often at low temperatures, poses practical challenges. 
Also, earlier works have primarily focused on creating and annihilating these topological spin textures by single femtosecond optical pulses \cite{Carbone2018,Shimojima2021,Fu2024,Zhu2018} or on topological objects which are confined to microstructured magnetic discs \cite{Zhu2018}.
Our experimental findings demonstrate that all-optical stochastic switching of N\'eel cap helicity occurs in dipolar skyrmions of \ce{Fe3Sn2} at room temperature and magnetic remanence. We analysed the light-induced switching behaviour of the N\'eel cap magnetization rotation of dipolar skyrmions and identified the underlying mechanism as a rapid temperature spike, due to the incident fs optical light pulse. The ensuing transient non-equilibrium state is followed by rapid cooling, leading to the re-emergence of the initial magnetic state but with a randomized helicity in the N\'eel surface spin twists. These experimental findings are corroborated by micromagnetic modelling and simulations of the electron phase shift map. The reliability of the switching process is evidenced by the preservation of the core Bloch state helicity across a broad range of incident optical fluences. The \ce{Fe3Sn2} lamella demonstrated long-term stability, with no visible structural damage evident after hundreds of light pulses. The initial results indicate the influence and potential controllability of the switching process by the micromagnetic environment of individual dipolar skyrmions and an external magnetic field. 
Moreover, the present study reports the interconversion of type I skyrmionic and type II magnetic bubbles in \ce{Fe3Sn2} upon excitation with fs optical pulses. 
Whereas, in the present experiments we have focused on the relaxed magnetic states after pulsed optical excitation, further interesting aspects are expected in the dynamical quenching processes at early delay times after excitation and for more complex optical fields, either structured temporally or in polarization degrees-of-freedom. Furthermore, \ce{Fe3Sn2} is an achiral material, thereby dipolar skyrmions without preferred handedness are generated. The interplay of structural, spin and optical chiralities in similar \emph{in situ} quenching-type experiments on chiral materials has the potential to provide a unique access to their coupling mechanisms. 
Our findings further establish the kagome-type \ce{Fe3Sn2} as an intriguing platform for the study of ultrafast optical control of compound topological spin textures and indicate a considerable potential for solutions to emerging technologies, such as probabilistic and neuromorphic computing \cite{Sutton2017,Borders2019,grollier_neuromorphic_2020,song_skyrmion-based_2020,hoffmann_quantum_2022}.

\section{Methods}\label{methods}
\textbf{TEM specimen preparation}
Focused Ga ion beam sputtering in a dual beam scanning electron microscope (ThermoFisher Helios NanoLab 460F1) was used to prepare a lamella from an \ce{Fe3Sn2} single crystal grown by a chemical transport method \cite{Kezsmarki2021} and to position the lamella on a copper Omniprobe support grid. The standard lift-off mechanism was followed to produce an electron transparent specimen with a thickness of approximately 200~nm. 

\textbf{Electron microscopy and optical setup}
To study the light-induced switching of magnetization textures in \ce{Fe3Sn2} dipolar skyrmions, we utilized the Regensburg UTEM, a modified JEOL JEM-F200 multi-purpose transmission electron microscope. This instrument is equipped with a photoemission-driven cold field-emitter electron gun and was, for the presented experiments, operated in the continuous electron beam mode at an electron energy of 200 keV.
We recorded electron micrographs in the low magnification mode of the TEM, with the objective lens turned off, with a nominal magnification of 8k and a defocus of 1.2~mm. The condenser system was adjusted for homogeneous illumination of the sample, with a spot size of 2 and a condenser system aperture with 100 $\mu$m diameter. Images were acquired by a complementary metal oxide semiconductor detector (TVIPS TemCam-XF416R, 4096 pixels × 4096 pixels, 15.5 $\mu$m pixel size, no binning, 1 s exposure time).
We optically pump the sample by means of an amplified Yb-doped potassium gadolinium tungstate (KGW) femtosecond laser system (Carbide, Light Conversion) and a collinear optical parametric amplifier (OPA, Orpheus HP, Light Conversion).
The laser system is operated at a repetition rate of 400~kHz. Out of the optical pulse train, individual pulses are sliced out for sample excitation using a computer-controlled Pockels cell. The pulse energy is controlled by a variable neutral density filter. For positioning the optical focus on the sample surface, a focusing lens mounted on a 3D piezo actuator, which is placed in front of a home-built light-incoupling unit at the height of the microscope pole piece.

\textbf{Micromagnetic simulation} 

The \ce{Fe3Sn2} sample was modelled as a uniaxial ferromagnet with the energy contributions from (i) ferromagnetic exchange, (ii) perpendicular uniaxial magnetocrystalline anisotropy, (iii) coupling to the external field, and (iv) long-range dipole-dipole interactions given by
\begin{equation}
E=A\sum_i\left(\nabla m_i\right)^2-K_1m_z^2-K_2m_z^4-\mu_0M_\mathrm{s}\mathbf{H}\cdot\mathbf{m}-\frac{1}{2}\mu_0M_\mathrm{s}\mathbf{H}_\mathrm{dip}\cdot \mathbf{m}
\end{equation}
where $E$ is the local energy density, $A=12\times 10^{-12}$ J/m is the exchange stiffness estimated from the Curie temperature, $\mathbf{m}=\mathbf{M}/M_\mathrm{s}$ is the local magnetization unit vector, $M_\mathrm{s}=5.66\times 10^5$ A/m is the saturation magnetization, $K_1=5.6\times 10^4$ J/m$^3$ and $K_2=6.5\times 10^3$ J/m$^3$ are the first- and second-order uniaxial anisotropy energy densities, extracted from magnetometry experiments on a single crystal. $\mathbf{H}$ is the external field vector, which includes the thermal field, and $\mathbf{H}_\mathrm{dip}$ is the magnetostatic field from long-range dipole-dipole interactions. The magnetization dynamics were computed in Mumax3 \cite{Mumax3} by numerically integrating the Landau-Lifshitz-Gilbert equation $\mathbf{\dot{m}}=-\gamma\left(\mathbf{m}\times\mathbf{B}_\mathrm{eff}\right)+\alpha\left(\mathbf{m}\times\mathbf{\dot{m}}\right)$, where $\gamma$ is the gyromagnetic ratio, and $\mathbf{B}_\mathrm{eff}=-\partial_\mathbf{m}E/M_\mathrm{s}$ is the effective field. 

The simulation cell size was set at 4~nm, with occasional tests with different cell sizes. The spin configuration of the sample was initialized with the magnetization along the perpendicular direction and then the LLG equation was integrated for 20 nanoseconds to reach a minimum energy state. Subsequently, a finite-temperature was introduced by means of a thermal field \cite{Mumax3}
\begin{equation}
B_\mathrm{T}=\mathbf{\eta}\sqrt{\frac{2\alpha k_\mathrm{B}T}{M_\mathrm{s}\gamma\Delta V \Delta t}}
\end{equation}
where $\mathbf{\eta}$ is a random vector that changes at every time step, $k_\mathrm{B}$ is the Boltzmann constant, $T$ is the temperature, $\Delta V$ is the volume of the simulation cell, and $\delta t$ is the time step. Notably, since micromagnetics consider macro-spins instead of elementary spins, the temperature response of the system becomes re-scaled and cell-size dependent, so that a significantly higher simulation temperature as compared to the experiments is required for partially melting the magnetic order. For these steps of the simulation, the integration time step was lowered to $10^{-15}$~s for numerical stability. The temperature was increased to $T_\mathrm{max}$ within 20~ps and then gradually decreased back to room temperature in 10 steps each with a duration of 10~ps.

The phase shift imparted on the imaging electron wave (shown in Fig.~\ref{fig1}e,f) was obtained by calculating the vector potential from the micromagnetic simulations. The magnetization vector $\mathbf{m}(x,y,z)$ was mapped onto a two-dimensional image by averaging for each value of $z$, and the average vector potential was computed with
\begin{equation}
    \mathbf{A}(r)=\frac{\mu_0}{4\pi}\int \frac{\mathbf{\tilde{m}}(r')\times \hat{\xi}}{\xi^2}dV'
\end{equation}
where the integration is over the 2D surface and $\mathbf{\xi}=\mathbf{r'}-\mathbf{r}$ is the separation between probing point and macro-spin position in the ($xy$)-plane. $\hat{\xi}$ denotes the unit vector in the direction of $\xi$. The phase map was constructed from the $z$ component of the averaged vector potential \cite{AharonovBohm1959} $\phi(x,y) =\frac{e\pi t M_\mathrm{s}}{\hbar}A_z(x,y)$, where $t$ is the thickness of the sample.

Comparison between phase shift maps of thin films with ideal structure vs. films with modified surface layers reveal that the characteristic central contrast in the dipolar skyrmion cores is the result of the increased and curled in-plane magnetization components of the N\'eel caps that are only obtained when one accounts for soft magnetic surfaces. 

\backmatter

\section*{Supplementary information}
Supplementary information available: \\
Supplementary Fig. S1: Segmented phase shift imparted on an electron beam by a dipolar skyrmion. \\
Supplementary Fig. S2: Type II magnetic bubbles in the micromagnetic simulations and Lorentz image simulations. \\ 
Section S1: Additional information on the statistical analysis of the light-induced switching behaviour of Néel caps in dipolar skyrmions with Fig. S3 and S4. \\
Movie S1. Reconstructed video, showing the light pulse induced switching of magnetization helicity in N\'eel caps of dipolar skyrmions in Fe$_3$Sn$_2$. \\
Movie S2. Reconstructed video, showing the light pulse induced transition from a type I to type II bubble.  \\
Movie S3. Reconstructed video, showing the dynamic behaviour of dipolar skyrmions in Fe$_3$Sn$_2$ upon applying a temperature spike, obtained from micromagnetic simulations. \\

\section*{Acknowledgments}

\par The authors are grateful for funding from the European Research Council under the European Union’s Horizon 2020 Research and Innovation Programme (Grant No. 856538, project ''3D MAGiC``), and from the Deutsche Forschungsgemeinschaft (DFG, German Research Foundation) (Project-ID 405553726, TRR270 and 49254781, TRR360 ). D.K. acknowledges the financial support to the “111” project (DB18015). 
Additionally, the authors would like to acknowledge financial support by the Volkswagen Foundation as part of the Lichtenberg Professorship “Ultrafast nanoscale dynamics probed by time-resolved electron imaging” and thank the DFG for funding the ultrafast transmission electron microscope (INST 184/211 1 FUGG). Furthermore, the authors acknowledge support by the Free State of Bavaria through the Lighthouse project “Free-electron states as ultrafast probes for qubit dynamics in solid-state platforms” within the Munich Quantum Valley initiative. Part of this work has been funded by the DFG through GRK 2905, project-ID 502572516.

\section*{Declarations}

The authors declare no competing interests. 

\newpage
\bibliography{FeSn}

\begin{thebibliography}{10}
\expandafter\ifx\csname url\endcsname\relax
  \def\url#1{\burl{#1}}\fi
\expandafter\ifx\csname urlprefix\endcsname\relax\def\urlprefix{URL }\fi
\providecommand{\bibinfo}[2]{#2}
\providecommand{\eprint}[2][]{\url{#2}}
\providecommand{\doi}[1]{\url{https://doi.org/#1}}
\bibcommenthead

\bibitem{Tokura2020}
\bibinfo{author}{Tokura, Y.} \& \bibinfo{author}{Kanazawa, N.}
\newblock \bibinfo{title}{Magnetic skyrmion materials}.
\newblock \emph{\bibinfo{journal}{Chemical Reviews}} \textbf{\bibinfo{volume}{121}}, \bibinfo{pages}{2857--2897} (\bibinfo{year}{2021}).

\bibitem{Zheng2018}
\bibinfo{author}{Zheng, F.} \emph{et~al.}
\newblock \bibinfo{title}{Experimental observation of chiral magnetic bobbers in {B}20-type {FeGe}}.
\newblock \emph{\bibinfo{journal}{Nature Nanotechnology}} \textbf{\bibinfo{volume}{13}}, \bibinfo{pages}{451--455} (\bibinfo{year}{2018}).

\bibitem{Kiselev2023}
\bibinfo{author}{Zheng, F.} \emph{et~al.}
\newblock \bibinfo{title}{Hopfion rings in a cubic chiral magnet}.
\newblock \emph{\bibinfo{journal}{Nature}} \textbf{\bibinfo{volume}{623}}, \bibinfo{pages}{718--723} (\bibinfo{year}{2023}).

\bibitem{kaiser_probabilistic_2021}
\bibinfo{author}{Kaiser, J.} \& \bibinfo{author}{Datta, S.}
\newblock \bibinfo{title}{Probabilistic computing with p-bits}.
\newblock \emph{\bibinfo{journal}{Applied Physics Letters}} \textbf{\bibinfo{volume}{119}}, \bibinfo{pages}{150503} (\bibinfo{year}{2021}).

\bibitem{muhlbauer_skyrmion_2009}
\bibinfo{author}{Mühlbauer, S.} \emph{et~al.}
\newblock \bibinfo{title}{Skyrmion {Lattice} in a {Chiral} {Magnet}}.
\newblock \emph{\bibinfo{journal}{Science}} \textbf{\bibinfo{volume}{323}}, \bibinfo{pages}{915--919} (\bibinfo{year}{2009}).

\bibitem{yu_near_2011}
\bibinfo{author}{Yu, X.~Z.} \emph{et~al.}
\newblock \bibinfo{title}{Near room-temperature formation of a skyrmion crystal in thin-films of the helimagnet {FeGe}}.
\newblock \emph{\bibinfo{journal}{Nature Materials}} \textbf{\bibinfo{volume}{10}}, \bibinfo{pages}{106--109} (\bibinfo{year}{2011}).

\bibitem{Fert2017}
\bibinfo{author}{Fert, A.}, \bibinfo{author}{Reyren, N.} \& \bibinfo{author}{Cros, V.}
\newblock \bibinfo{title}{Magnetic skyrmions: advances in physics and potential applications}.
\newblock \emph{\bibinfo{journal}{Nature Reviews Materials}} \textbf{\bibinfo{volume}{2}}, \bibinfo{pages}{17031} (\bibinfo{year}{2017}).

\bibitem{Du2022}
\bibinfo{author}{Wang, W.} \emph{et~al.}
\newblock \bibinfo{title}{Electrical manipulation of skyrmions in a chiral magnet}.
\newblock \emph{\bibinfo{journal}{Nature Communications}} \textbf{\bibinfo{volume}{13}}, \bibinfo{pages}{1593} (\bibinfo{year}{2022}).

\bibitem{Fenner2009}
\bibinfo{author}{Fenner, L.~A.}, \bibinfo{author}{Dee, A.~A.} \& \bibinfo{author}{Wills, A.~S.}
\newblock \bibinfo{title}{Non-collinearity and spin frustration in the itinerant kagome ferromagnet {F}e$_3${S}n$_2$}.
\newblock \emph{\bibinfo{journal}{Journal of Physics: Condensed Matter}} \textbf{\bibinfo{volume}{21}}, \bibinfo{pages}{452202} (\bibinfo{year}{2009}).

\bibitem{Kida_2011}
\bibinfo{author}{Kida, T.} \emph{et~al.}
\newblock \bibinfo{title}{The giant anomalous {H}all effect in the ferromagnet {F}e$_3${S}n$_2$ frustrated kagome metal}.
\newblock \emph{\bibinfo{journal}{Journal of Physics: Condensed Matter}} \textbf{\bibinfo{volume}{23}}, \bibinfo{pages}{112205} (\bibinfo{year}{2011}).

\bibitem{Schilbert2022}
\bibinfo{author}{Schilberth, F.} \emph{et~al.}
\newblock \bibinfo{title}{Magneto-optical detection of topological contributions to the anomalous {H}all effect in a kagome ferromagnet}.
\newblock \emph{\bibinfo{journal}{Phys. Rev. B}} \textbf{\bibinfo{volume}{106}}, \bibinfo{pages}{144404} (\bibinfo{year}{2022}).

\bibitem{Lin2018}
\bibinfo{author}{Lin, Z.} \emph{et~al.}
\newblock \bibinfo{title}{Flatbands and emergent ferromagnetic ordering in {F}e$_3${S}n$_2$ kagome lattices}.
\newblock \emph{\bibinfo{journal}{Phys. Rev. Lett.}} \textbf{\bibinfo{volume}{121}}, \bibinfo{pages}{096401} (\bibinfo{year}{2018}).

\bibitem{Ye2018}
\bibinfo{author}{Ye, L.} \emph{et~al.}
\newblock \bibinfo{title}{Massive {D}irac fermions in a ferromagnetic kagome metal}.
\newblock \emph{\bibinfo{journal}{Nature}} \textbf{\bibinfo{volume}{555}}, \bibinfo{pages}{638--642} (\bibinfo{year}{2018}).

\bibitem{Kang2020}
\bibinfo{author}{Kang, M.} \emph{et~al.}
\newblock \bibinfo{title}{Dirac fermions and flat bands in the ideal kagome metal {F}e{S}n}.
\newblock \emph{\bibinfo{journal}{Nature Materials}} \textbf{\bibinfo{volume}{19}}, \bibinfo{pages}{163--169} (\bibinfo{year}{2020}).

\bibitem{Kezsmarki2021}
\bibinfo{author}{Altthaler, M.} \emph{et~al.}
\newblock \bibinfo{title}{Magnetic and geometric control of spin textures in the itinerant kagome magnet {F}e$_3${S}n$_2$}.
\newblock \emph{\bibinfo{journal}{Phys. Rev. Res.}} \textbf{\bibinfo{volume}{3}}, \bibinfo{pages}{043191} (\bibinfo{year}{2021}).

\bibitem{Kong2023b}
\bibinfo{author}{Kong, L.} \emph{et~al.}
\newblock \bibinfo{title}{Observation of hybrid magnetic skyrmion bubbles in ${\mathrm{fe}}_{3}{\mathrm{sn}}_{2}$ nanodisks}.
\newblock \emph{\bibinfo{journal}{Phys. Rev. B}} \textbf{\bibinfo{volume}{107}}, \bibinfo{pages}{174425} (\bibinfo{year}{2023}).

\bibitem{Du2024}
\bibinfo{author}{Kong, L.} \emph{et~al.}
\newblock \bibinfo{title}{Diverse helicities of dipolar skyrmions}.
\newblock \emph{\bibinfo{journal}{Phys. Rev. B}} \textbf{\bibinfo{volume}{109}}, \bibinfo{pages}{014401} (\bibinfo{year}{2024}).

\bibitem{Du2023}
\bibinfo{author}{Tang, J.} \emph{et~al.}
\newblock \bibinfo{title}{Combined magnetic imaging and anisotropic magnetoresistance detection of dipolar skyrmions}.
\newblock \emph{\bibinfo{journal}{Advanced Functional Materials}} \textbf{\bibinfo{volume}{33}}, \bibinfo{pages}{2207770} (\bibinfo{year}{2023}).

\bibitem{klaui_perspective_2018}
\bibinfo{author}{Everschor-Sitte, K.}, \bibinfo{author}{Masell, J.}, \bibinfo{author}{Reeve, R.~M.} \& \bibinfo{author}{Kläui, M.}
\newblock \bibinfo{title}{Perspective: {Magnetic} skyrmions—{Overview} of recent progress in an active research field}.
\newblock \emph{\bibinfo{journal}{Journal of Applied Physics}} \textbf{\bibinfo{volume}{124}}, \bibinfo{pages}{240901} (\bibinfo{year}{2018}).

\bibitem{Rasing2025}
\bibinfo{author}{Strungaru, M.} \emph{et~al.}
\newblock \bibinfo{title}{All optical switching: a path to recording technology beyond hamr}.
\newblock \emph{\bibinfo{journal}{IEEE Transactions on Magnetics}} \textbf{\bibinfo{volume}{-}}, \bibinfo{pages}{--} (\bibinfo{year}{2025}).

\bibitem{Lambert2014}
\bibinfo{author}{Lambert, C.-H.} \emph{et~al.}
\newblock \bibinfo{title}{All-optical control of ferromagnetic thin films and nanostructures}.
\newblock \emph{\bibinfo{journal}{Science}} \textbf{\bibinfo{volume}{345}}, \bibinfo{pages}{1337--1340} (\bibinfo{year}{2014}).

\bibitem{beaurepaire_ultrafast_1996}
\bibinfo{author}{Beaurepaire, E.}, \bibinfo{author}{Merle, J.-C.}, \bibinfo{author}{Daunois, A.} \& \bibinfo{author}{Bigot, J.-Y.}
\newblock \bibinfo{title}{Ultrafast {Spin} {Dynamics} in {Ferromagnetic} {Nickel}}.
\newblock \emph{\bibinfo{journal}{Physical Review Letters}} \textbf{\bibinfo{volume}{76}}, \bibinfo{pages}{4250--4253} (\bibinfo{year}{1996}).

\bibitem{Rasing2010}
\bibinfo{author}{Kirilyuk, A.}, \bibinfo{author}{Kimel, A.~V.} \& \bibinfo{author}{Rasing, T.}
\newblock \bibinfo{title}{Ultrafast optical manipulation of magnetic order}.
\newblock \emph{\bibinfo{journal}{Reviews of Modern Physics}} \textbf{\bibinfo{volume}{82}}, \bibinfo{pages}{2731} (\bibinfo{year}{2010}).

\bibitem{Schafer2017}
\bibinfo{author}{Eggebrecht, T.} \emph{et~al.}
\newblock \bibinfo{title}{Light-induced metastable magnetic texture uncovered by in situ {L}orentz microscopy}.
\newblock \emph{\bibinfo{journal}{Phys. Rev. Lett.}} \textbf{\bibinfo{volume}{118}}, \bibinfo{pages}{097203} (\bibinfo{year}{2017}).

\bibitem{Adam2025}
\bibinfo{author}{Chen, X.} \emph{et~al.}
\newblock \bibinfo{title}{Ultrafast demagnetization in ferromagnetic materials: Origins and progress}.
\newblock \emph{\bibinfo{journal}{Physics Reports}} \textbf{\bibinfo{volume}{1102}}, \bibinfo{pages}{1--63} (\bibinfo{year}{2025}).
\newblock \bibinfo{note}{Ultrafast demagnetization in ferromagnetic materials: Origins and progress}.

\bibitem{stanciu_all-optical_2007}
\bibinfo{author}{Stanciu, C.~D.} \emph{et~al.}
\newblock \bibinfo{title}{All-{Optical} {Magnetic} {Recording} with {Circularly} {Polarized} {Light}}.
\newblock \emph{\bibinfo{journal}{Physical Review Letters}} \textbf{\bibinfo{volume}{99}}, \bibinfo{pages}{047601} (\bibinfo{year}{2007}).

\bibitem{radu_transient_2011}
\bibinfo{author}{Radu, I.} \emph{et~al.}
\newblock \bibinfo{title}{Transient ferromagnetic-like state mediating ultrafast reversal of antiferromagnetically coupled spins}.
\newblock \emph{\bibinfo{journal}{Nature}} \textbf{\bibinfo{volume}{472}}, \bibinfo{pages}{205--208} (\bibinfo{year}{2011}).

\bibitem{park_4d_2010}
\bibinfo{author}{Park, H.~S.}, \bibinfo{author}{Baskin, J.~S.} \& \bibinfo{author}{Zewail, A.~H.}
\newblock \bibinfo{title}{{4D} {Lorentz} {Electron} {Microscopy} {Imaging}: {Magnetic} {Domain} {Wall} {Nucleation}, {Reversal}, and {Wave} {Velocity}}.
\newblock \emph{\bibinfo{journal}{Nano Letters}} \textbf{\bibinfo{volume}{10}}, \bibinfo{pages}{3796--3803} (\bibinfo{year}{2010}).

\bibitem{schliep_picosecond_2017}
\bibinfo{author}{Schliep, K.~B.}, \bibinfo{author}{Quarterman, P.}, \bibinfo{author}{Wang, J.-P.} \& \bibinfo{author}{Flannigan, D.~J.}
\newblock \bibinfo{title}{Picosecond {Fresnel} transmission electron microscopy}.
\newblock \emph{\bibinfo{journal}{Applied Physics Letters}} \textbf{\bibinfo{volume}{110}}, \bibinfo{pages}{222404} (\bibinfo{year}{2017}).

\bibitem{Schafer2018}
\bibinfo{author}{Rubiano~da Silva, N.} \emph{et~al.}
\newblock \bibinfo{title}{Nanoscale mapping of ultrafast magnetization dynamics with femtosecond {L}orentz microscopy}.
\newblock \emph{\bibinfo{journal}{Phys. Rev. X}} \textbf{\bibinfo{volume}{8}}, \bibinfo{pages}{031052} (\bibinfo{year}{2018}).

\bibitem{Carbone2018}
\bibinfo{author}{Berruto, G.} \emph{et~al.}
\newblock \bibinfo{title}{Laser-induced skyrmion writing and erasing in an ultrafast cryo-lorentz transmission electron microscope}.
\newblock \emph{\bibinfo{journal}{Phys. Rev. Lett.}} \textbf{\bibinfo{volume}{120}}, \bibinfo{pages}{117201} (\bibinfo{year}{2018}).

\bibitem{Ropers2020}
\bibinfo{author}{M{\"o}ller, M.}, \bibinfo{author}{Gaida, J.~H.}, \bibinfo{author}{Sch{\"a}fer, S.} \& \bibinfo{author}{Ropers, C.}
\newblock \bibinfo{title}{Few-nm tracking of current-driven magnetic vortex orbits using ultrafast {L}orentz microscopy}.
\newblock \emph{\bibinfo{journal}{Communications Physics}} \textbf{\bibinfo{volume}{3}}, \bibinfo{pages}{36} (\bibinfo{year}{2020}).

\bibitem{Zhu2024}
\bibinfo{author}{Liu, C.}, \bibinfo{author}{Reisbick, S.} \& \bibinfo{author}{Zhu, Y.}
\newblock \bibinfo{title}{Capturing spin waves with microwave-mediated stroboscopic electron microscopy}.
\newblock \emph{\bibinfo{journal}{Microscopy and Microanalysis}} \textbf{\bibinfo{volume}{30}}, \bibinfo{pages}{ozae044.709} (\bibinfo{year}{2024}).

\bibitem{liu_correlated_2025}
\bibinfo{author}{Liu, C.} \emph{et~al.}
\newblock \bibinfo{title}{Correlated spin-wave generation and domain-wall oscillation in a topologically textured magnetic film}.
\newblock \emph{\bibinfo{journal}{Nature Materials}}  (\bibinfo{year}{2025}).

\bibitem{fan_spatiotemporal_2025}
\bibinfo{author}{Fan, Y.}, \bibinfo{author}{Cao, G.}, \bibinfo{author}{Jiang, S.}, \bibinfo{author}{Åkerman, J.} \& \bibinfo{author}{Weissenrieder, J.}
\newblock \bibinfo{title}{Spatiotemporal observation of surface plasmon polariton mediated ultrafast demagnetization}.
\newblock \emph{\bibinfo{journal}{Nature Communications}} \textbf{\bibinfo{volume}{16}}, \bibinfo{pages}{873} (\bibinfo{year}{2025}).

\bibitem{Zhu2018}
\bibinfo{author}{Fu, X.} \emph{et~al.}
\newblock \bibinfo{title}{Optical manipulation of magnetic vortices visualized in situ by {L}orentz electron microscopy}.
\newblock \emph{\bibinfo{journal}{Science Advances}} \textbf{\bibinfo{volume}{4}}, \bibinfo{pages}{eaat3077} (\bibinfo{year}{2018}).

\bibitem{Fu2024}
\bibinfo{author}{Li, Z.} \emph{et~al.}
\newblock \bibinfo{title}{Room-temperature sub-100 nm {N\'e}el-type skyrmions in non-stoichiometric van der {W}aals ferromagnet {F}e$_{3-x}${G}a{T}e$_2$ with ultrafast laser writability}.
\newblock \emph{\bibinfo{journal}{Nature Communications}} \textbf{\bibinfo{volume}{15}}, \bibinfo{pages}{1017} (\bibinfo{year}{2024}).

\bibitem{Shimojima2021}
\bibinfo{author}{Shimojima, T.} \emph{et~al.}
\newblock \bibinfo{title}{Nano-to-micro spatiotemporal imaging of magnetic skyrmion’s life cycle}.
\newblock \emph{\bibinfo{journal}{Science Advances}} \textbf{\bibinfo{volume}{7}}, \bibinfo{pages}{eabg1322} (\bibinfo{year}{2021}).

\bibitem{Carbone2022}
\bibinfo{author}{Tengdin, P.} \emph{et~al.}
\newblock \bibinfo{title}{Imaging the ultrafast coherent control of a skyrmion crystal}.
\newblock \emph{\bibinfo{journal}{Phys. Rev. X}} \textbf{\bibinfo{volume}{12}}, \bibinfo{pages}{041030} (\bibinfo{year}{2022}).

\bibitem{tang_two-dimensional_2021}
\bibinfo{author}{Tang, J.} \emph{et~al.}
\newblock \bibinfo{title}{Two-dimensional characterization of three-dimensional magnetic bubbles in {Fe$_3$Sn$_2$} nanostructures}.
\newblock \emph{\bibinfo{journal}{National Science Review}} \textbf{\bibinfo{volume}{8}}, \bibinfo{pages}{nwaa200} (\bibinfo{year}{2021}).

\bibitem{Sutton2017}
\bibinfo{author}{Sutton, B.}, \bibinfo{author}{Camsari, K.~Y.}, \bibinfo{author}{Behin-Aein, B.} \& \bibinfo{author}{Datta, S.}
\newblock \bibinfo{title}{Intrinsic optimization using stochastic nanomagnets}.
\newblock \emph{\bibinfo{journal}{Scientific Reports}} \textbf{\bibinfo{volume}{7}}, \bibinfo{pages}{44370} (\bibinfo{year}{2017}).

\bibitem{Borders2019}
\bibinfo{author}{A.Borders, W.} \emph{et~al.}
\newblock \bibinfo{title}{Integer factorization using stochastic magnetic tunnel junctions}.
\newblock \emph{\bibinfo{journal}{Nature}} \textbf{\bibinfo{volume}{573}}, \bibinfo{pages}{390--393} (\bibinfo{year}{2019}).

\bibitem{grollier_neuromorphic_2020}
\bibinfo{author}{Grollier, J.} \emph{et~al.}
\newblock \bibinfo{title}{Neuromorphic spintronics}.
\newblock \emph{\bibinfo{journal}{Nature Electronics}} \textbf{\bibinfo{volume}{3}}, \bibinfo{pages}{360--370} (\bibinfo{year}{2020}).

\bibitem{song_skyrmion-based_2020}
\bibinfo{author}{Song, K.~M.} \emph{et~al.}
\newblock \bibinfo{title}{Skyrmion-based artificial synapses for neuromorphic computing}.
\newblock \emph{\bibinfo{journal}{Nature Electronics}} \textbf{\bibinfo{volume}{3}}, \bibinfo{pages}{148--155} (\bibinfo{year}{2020}).

\bibitem{hoffmann_quantum_2022}
\bibinfo{author}{Hoffmann, A.} \emph{et~al.}
\newblock \bibinfo{title}{Quantum materials for energy-efficient neuromorphic computing: {Opportunities} and challenges}.
\newblock \emph{\bibinfo{journal}{APL Materials}} \textbf{\bibinfo{volume}{10}}, \bibinfo{pages}{070904} (\bibinfo{year}{2022}).

\bibitem{Mumax3}
\bibinfo{author}{Vansteenkiste, A.} \emph{et~al.}
\newblock \bibinfo{title}{The design and verification of {M}u{M}ax3}.
\newblock \emph{\bibinfo{journal}{AIP Advances}} \textbf{\bibinfo{volume}{4}}, \bibinfo{pages}{107133} (\bibinfo{year}{2014}).

\bibitem{AharonovBohm1959}
\bibinfo{author}{Aharonov, Y.} \& \bibinfo{author}{Bohm, D.}
\newblock \bibinfo{title}{Significance of electromagnetic potentials in the quantum theory}.
\newblock \emph{\bibinfo{journal}{Phys. Rev.}} \textbf{\bibinfo{volume}{115}}, \bibinfo{pages}{485--491} (\bibinfo{year}{1959}).

\end{thebibliography}

\end{document}